\documentstyle[prl,aps,graphicx]{revtex}
\begin{document}
\twocolumn[
\hsize\textwidth\columnwidth\hsize\csname@twocolumnfalse\endcsname

\title{The theory of magneto-transport in quantum dots: 3D-0D and 2D-0D tunnelling and selection rules for the angular momentum.}
\author{B.Jouault$^{1}$, M. Boero$^{2}$, G.Faini~\footnote{}$^{1}$, J.C. Inkson$^{2}$}
\address{
$^{1}$ L2M-CNRS, 196 Avenue H. Rav\'era, BP107, 92225 Bagneux Cedex, France \\
$^{2}$ Department of Physics, University of Exeter, Stocker Road, Exeter,EX4 4QL, 
UK \\
$^{*}$e-mail:giancarlo.faini@L2M.CNRS.fr}
\maketitle
\begin{abstract}

A study of magneto-transport through quantum dots is presented. 
The model allows to analyze tunnelling both from bulk-like contacts and from 2D 
accumulation
layers. The fine features in the I-V characteristics due to the quantum dot states
are known to be shifted to different voltages depending upon the value of the 
magnetic
field. While this effect is also well reproduced by our calculations, in this work 
we 
concentrate on the amplitude of each current resonance as a function of the magnetic 
field. Such amplitudes show oscillations reflecting the variation of the density 
of states at the Fermi energy 
in the emitter. Furthermore the amplitude increases as a function of the magnetic 
field for certain features while it decreases for others. In particular we 
demonstrate that the behaviour of the amplitude of the current resonances is linked to 
the value of the angular momentum of each dot level through which tunnelling 
occurs. We show that a selection rule on the angular momentum must be satisfied. 
As a consequence, tunnelling through specific dot states is strongly suppresses 
and sometimes prohibited altogether by the presence of the magnetic field. This 
will allow to extract from the experimental curves detailed information on the 
nature of the quantum dot wavefunctions involved in the electronic transport. 
Furthermore, when tunnelling occurs from a 2D accumulation layer to the quantum 
dot, the presence of a magnetic field hugely increases the strength of some 
resonant features. This effect is predicted by our model and, to the best of our 
knowledge, has never been observed.
\bigskip
\end{abstract}


]
\newcommand{\MASS}{m^{*}}
\bigskip

\section{Introduction}
 The manufacturing of structures where the electronic motion is confined in 1,2,3 
directions has been possible for several years and a large amount of experimental 
evidence has been collected to study the behaviour of electrons in 2,1,0 
dimensions. This has resulted in a series of new physical phenomena and has raised 
hopes of building devices where the behaviour of single electrons is controlled.
Quantum dots are structures where the electron motion is confined in all 
directions and therefore their energy spectrum is discrete. Several spectroscopic 
techniques such as 
linear and non-linear transport, capacitance measurements and far-infrared 
spectroscopy have been used to investigate the energy spectra of quantum dots. The 
addition of an external magnetic field has proved to be an excellent way to probe 
the low-dimensional systems since it introduces a structure-independent 
quantisation that is superimposed to that induced by the fabrication process. 
Quantum dots obtained from double-barrier resonant tunnelling semiconductor 
heterostructures are now used by several groups for magneto-tunnelling 
investigations~\cite{ree89,ram91,bo92,kou97}, since it is possible to probe the 
quantum dot spectra in several regimes from the very low to the very high electron 
number occupancy, tuning the relative strength between confinement and 
electron-electron interactions. The spectroscopic investigation is still exciting 
and far from complete, and several issues such as the importance of the 
electrostatic interaction among the carriers, detailed characteristics of the 
quantum dot spectra, degeneracy of the states and 
spin of the different multiplets must still be fully solved.

 Practically, in all the previous studies on non-linear transport, the emphasis was 
placed on the displacement of the I-V features produced in the quantum dot states 
as a function of the magnetic field in order to probe the quantum dot 
spectrum~\cite{ree89,ram91,bo92,kou97}.
 On a more fundamental point lies the issue of actually probing the confined 
wavefunctions inside a quantum dot. This analysis has been carried out by P.H. 
Beton {\it et al.}~\cite{bet95} for the states in a quantum wire. The method 
consists in measuring the current voltage characteristics of a quantum wire in the 
presence of a magnetic field as in the previous studies, but to focus the analysis 
on the magnitude of the current of each peak. Indeed, it is shown that the 
amplitude of the current can be 
associated with the Fourier transform of the electron wave function in the 
direction across the wire.
Very recently~\cite{jou98}, we reported on a similar analysis of the experimental 
data collected on quantum dots fabricated from a double-barrier resonant 
tunnelling structure by an ion implantation and electrode regrowth 
process~\cite{fai96}.
To the best of our knowledge this was the first time such an analysis had been 
performed for quantum
dots. 

 In this work we lay the theoretical foundations to obtain a detailed
understanding of the electronic properties of quantum dots from magneto-transport 
experiments. 
In particular we address
the problem of non-linear magneto-transport in 3D-0D-3D and 2D-0D-2D structures. 
This can yield
informations not only on the energy position of the dot states, but also on the 
actual
properties of the wavefunctions, i.e their angular momentum and their spatial 
extension. Extending the work performed by P.H. Beton {\it et al.}, we find
that also for quantum dots the height of the resonances in the current-voltage
characteristics depends strongly on the magnetic field. The height of
each resonance shows oscillations as a function of the magnetic field which reflect the
oscillations in the density of states at the Fermi energy. Furthermore the 
behaviour of
the amplitude of the fine features changes considerably depending on the particular 
quantum dot
state considered. It is found that a general trend is obeyed depending on the 
value of
the angular momentum of the dot state, and that the angular momentum introduces a 
selection rule that strongly suppresses tunnelling through states with angular 
momentum co-linear to the magnetic field. Above certain values of the magnetic 
field the 
selection rules totally prohibits tunnelling through such states.
The last section deals with the 2D-0D magneto-transport occurring in those 
experimental structures where an accumulation layer is formed in the emitter. Here 
the effect of the magnetic field is totally different 
from the 3D-0D-3D case. We show in particular that the amplitude of the current 
resonances are a direct measure of the overlap between the Landau levels in the emitter 
and the dot states involved in the tunnelling process. The presence of the 
magnetic field hugely increases the tunnelling current flowing through the quantum 
dot ground state and indeed through all the states with principal quantum number 
n=0 and non-positive angular momentum. It must be noted that this effect applies 
to the quantum dot ground state, i.e. when only a single electron occupies the 
dot. Therefore it remains valid even in the presence of electron-electron 
repulsion inside the dot. 
 The theoretical analysis developed here will help to distinguish the nature of 
the various features in the experimental investigations~\cite{jou98}, to establish 
unambiguously the dimensionality of the electrodes and to gain further knowledge 
of the properties of quantum dots.

\section{General remarks}
 The current-voltage characteristics of quantum dot structures are dominated by 
the
concept of resonant tunnelling. The current flows through a specific quantum dot 
state
only if its energy falls within the energy range of the occupied states in the 
emitter
conduction band. The total current flowing through the structure for a given 
applied
voltage bias is the sum of the current flowing through each quantum dot state. In 
systems such as double-barrier resonant tunnelling heterostructures, at zero 
voltage no current flows through the structure because the quantum dot states lie 
above the Fermi energy in the emitter. As the voltage is raised, the quantum dot 
states fall below the Fermi energy, resonant tunnelling becomes possible and the 
current increases. It follows that the voltage at which each dot state becomes 
available for tunnelling is associated with the energy of the dot state itself. 
Therefore transport measurements provide informations on the quantum dot spectrum.

 It has long been argued that transport measurements must be performed in the 
presence of a magnetic
field in order to be a powerful spectroscopic tool. This is because the presence 
of impurity states in the quantum dot region can produce features in the I-V 
characteristics that are similar to the ones produced by the quantum dot states. 
The presence of a magnetic field allows to alter the geometry and the strength of 
the confinement of the quantum dot in an controllable fashion so that the 
properties of the quantum dot can be differentiated from those due to impurities. 
Furthermore the features produced by the quantum dot can be analyzed in much 
greater details by the application of a magnetic field, hence the importance of 
understanding in details the physics of magneto-tunnelling in quantum dots.

 The typical structure to study vertical transport in quantum dots consists of a 
double-barrier quantum well which is confined in the lateral directions. The 
lateral confinement can be introduced by means of several methods such as 
etching~\cite{ree89,ram91,bo92,bet95}, ion-bombardment~\cite{fai96}, and the 
presence of gates~\cite{kou97}. For all these methods the resulting confining 
potential can be very accurately approximated by a parabola. In the growth 
direction the heterostructure confining potential V(z) is much stronger than the 
lateral one and has the shape of a square well. Therefore the Hamiltonian for the 
dot containing N particles can be written as:

\begin{equation}
H=\sum_{i=1}^{N}\frac{(\vec{p_{i}}-e\vec{A_{i}})^{2}}{2m^{*}}+\frac{1}{2}m^{*}
\omega_{0}^{2}
\sum_{i=1}^{N}
(x_{i}^{2}+y_{i}^{2})+V(z)
\end{equation}

In this work we are going to analyze the case where the magnetic field is applied 
along
the vertical direction of the structure, i.e. along the direction of the current. 
As we are going to deal with small structures where the single electron effects 
dominate over the electron-electron repulsion~\cite{boe94}, we do not take into 
account the Coulomb repulsion inside the dot. This approximation considerably 
simplifies the problem of transport through quantum dots in the non-linear regime 
and, while it is not theoretically satisfactory, it has proven correct in several 
circumstances both experimentally~\cite{ree89,ram91,bo92,bet95,jou98,ree88} and theoretically~\cite{boe94,bry89,bry91}. 
Nevertheless, the analysis presented here is valid also in the presence of the 
electron electron interaction when this can be approximated with the Hubbard-like hamiltonian. This has been proved a valid approach for small dots of radius ~50 nm or less~\cite{boe94}, as those ones considered here.
While many theoretical efforts, based on non-equilibrium Green's functions, have been 
devoted to the development of a theory of transport in the non-linear regime for 
correlated systems~\cite{meir92}, a comprehensive theory capable of 
dealing with a multi-electron system and including partial occupancy and 
non-equilibrium filling of the dot states in a quantitative fashion does not 
exist.
In the single particle approximation eq.(1) can be solved exactly. Thus the single 
electron spectrum, neglecting the effects of spin, is given by:

\begin{equation}
E_{n_{d},m_{d}}=\hbar\omega_{d}(2n_{d}+|m_{d}|+1)+\frac{1}{2}\hbar\omega_{c}m_{d}+ 
E_{0}
\end{equation}

where $\omega_{d}=\sqrt{\omega_{0}^{2}+\frac{\omega_{c}^{2}}{4}}$ , $\omega_{0}$
characterizes the strength of the lateral confinement in the dot, $\omega_{c}$ is 
the cyclotron
frequency and $E_{0}$ is the energy of the level due to the confinement in the 
growth
direction. The principal quantum number $n_{d}$ can take any non-negative integer
value. On the contrary, the quantum number $m_{d}$, associated with the angular 
momentum in the
direction of the current and the magnetic field, can take any positive or negative
integer value. The eigenfunctions related to the eq. (2) are given by:

\begin{equation}
\Phi_{d}(r,\varphi,z)=\frac{e^{im_{d}\varphi}}{\sqrt{2\pi}}R_{n_{d},m_{d}}(r)\Psi_
{d}(z)
\end{equation}
where $R_{n_{d},m_{d}}(r)$ is a function that can be expressed in terms of the
hypergeometric function \cite{lan,foc28}:
\begin{equation}
R_{n,m}(r) \propto e^{-\xi/2} \xi^{|m|/2} F(-n,|m|+1,\xi)
\end{equation}
where $\xi=\omega_{d}m^{*}r^{2}/\hbar$.
$\Psi_{d}(z)$ is the wavefunction in the growth direction which consists of the 
familiar ground state of a square well of finite height.

As pioneered by Bardeen~\cite{bar61} and further developed by Payne~\cite{pay86} 
and by Liu and Aers~\cite{liu88}, the tunnelling current can be calculated in the 
framework of a transfer matrix approach. The system is divided into three 
sub-systems, the two contacts and the quantum dot, and the spectrum for each 
sub-system considered isolated is determined.

The tunnelling process is modeled as a two-step process in which electrons first 
tunnel from the emitter to the dot and eventually from the dot to the collector. 
In spite of its sequential nature, this approach is well-known to reproduce the 
correct physics even in the resonant tunnelling regime~\cite{che93}.
The current for each step can be calculated from the knowledge of the 
wavefunctions in the three different regions. At equilibrium the current flowing 
from the emitter to the dot must be equal to the one going from the dot to the 
collector. Imposing this condition, enables one to calculate the current as well 
as the 
occupancy of the dot states. The strength of the tunnelling current flowing 
through each dot state is determined by
the overlap integral between the dot wavefunctions and those of the 
contacts~\cite{pay86,liu88,boe94,boe97}. Following Bardeen, this is given by:
\begin{equation}
M_{e \rightarrow d} = \left( \frac {\hbar^{2}} {2\MASS} \right) \int_{s} (
\Phi_{e}\nabla\Phi_{d}^{*} - \Phi_{d}^{*}\nabla\Phi_{e}).d\mathbf{S}
\end{equation}
for the transition of the electrons from the emitter to the dot, here S is any 
surface across the emitter-dot barrier. A similar formula 
applies for the tunnelling between the dot and the collector.

 In order to calculate the current it is necessary to know not only the electron 
wave functions in the dot, but also those of the contacts. As these depend on the 
actual shape of the contacts, we are going to discuss two different cases 
separately.

\section{First case: 3D contacts}
Let us first suppose that no geometrical confinement is present in the contacts. 
This is indeed the case in many experimental 
situations~\cite{bo92,kou97,jou98,fai96} where dots are either obtain by means of 
deep etching or by ion bombardment.
In the presence of an applied
magnetic field, the energy spectrum in the contacts takes the form of highly 
degenerate
1D Landau sub-bands which are occupied up to the Fermi energy. The energy spectrum 
is
given by:
\begin{equation}
E_{n_{e},m_{e},k_{z}} =
\hbar\omega_{c}(n_{e}+\frac{m_{e}+|m_{e}|}{2}+\frac{1}{2})+\frac{\hbar^{2}k^{2}_{z
}}{2m^{*}}
\end{equation}
where $k_{z}$ is the momentum in the growth direction. As for the quantum dot, 
$m_{e}$ can take any integer value while $n_{e}$ can take all non-negative integer 
values. Note however the difference between positive and negative values of 
$m_{e}$. While each Landau sub-bands contains all negative values of $m_{e}$, the 
lowest sub-band, of minimum energy $\frac{1}{2}\hbar\omega_{c}$, does not contain 
positive values of $m_{e}$, the first excited sub-band contains only  $m_{e}=1$, 
the second excited sub-band  $m_{e}=1,2$ and so on.. In general terms the maximum 
value 
of the angular momentum in the j-th sub-band is $m_{e}=j-1$.  

The eigenfunctions are plane waves along the vertical growth direction while for 
the lateral part they are of the same shape as those of the quantum dot:
\begin{equation}
\Phi_{e}(r,\varphi,z)=\frac{e^{im_{e}\varphi}}{\sqrt{2\pi}}R_{n_{e},m_{e}}(r)\Psi_
{e}(z)
\label{eq:phie}
\end{equation}
where $\Psi_{e}(z)$ is the wavefunction defined by the one dimensional square 
potential:
$V(z)=0$ before the emitter barrier for $z \leq b$ and $V(z)=V_{0}$ for $z \geq 
b$, b being the 
abscissa of the left side of the barrier and $V_{0}$ the height of the barrier. In 
this case $\Psi_e(z)$ is given by:
\begin{equation}
\Psi_{e}=\left\{
\matrix{
\sqrt{\frac{2}{L_{e}}}\sin(k_{e}[z-b+L_{e}]) & z \leq b \cr
\sqrt{\frac{2}{L_{e}}}\sin(k_{e}L_{e}) \exp( -K_{e}[z-b] ) & z \geq b \cr
}\right.
\label{eq:psi}
\end{equation}
where $L_{e}$ is a normalization constant, $k_{e}= \sqrt{2m^{*}E}/\hbar$ is the 
lateral momentum and $K_{e}$ is defined as $K_{e}= \sqrt{2m^{*}(V_{0}-E)}/\hbar$. 
E is the energy in the vertical direction.
The spectrum and wavefunctions of the collector are exactly of the same shape as 
those
of the emitter.

The overlap between the quantum dot states and those in the contacts depends on 
the
height and width of the barriers along the direction through which the current 
flows. In the
lateral directions it depends on the particular quantum dot and contact states
considered. The overlap integral for a one dimensional problem with barriers of 
square
shape has been studied in details by Payne~\cite{pay86}. In this work we will 
adopt
the same approach to treat the overlap in the growth direction and we will extend 
the
method to the three dimensional case. In order to do so one must calculate the 
overlap
integrals in the lateral directions between the quantum dot and the contact 
states. These are of the form:
\begin{eqnarray}
\label{eq:over}
M_{n_{d},m_{d}:n_{e},m_{e}}^{\|}= \\ \nonumber
\int^{2\pi}_{0} \! \! \! \! 
e^{-i(m_{d}-m_{e})\varphi}
d\varphi\int_{0}^{\infty} \! \! \! \! rR_{n_{d},m_{d}}(r)
R_{n_{e},m_{e}}(r)dr
\end{eqnarray}
where the e and d refer to states in the contacts and in the dot respectively. 
The overlap integral~\ref{eq:over} can be calculated analytically for all values 
of $m_{d},n_{d},m_{e}, n_{e}$. Furthermore it is clear from~\ref{eq:over} that 
only the states with $m_{d}=m_{e}=m$ produce a non zero overlap. Therefore from 
the highly degenerate Landau sub-bands in the contacts, at most only one 
particular 
state will contribute to the tunnelling current flowing through each dot state: 
the one with the right value of the angular momentum $m_{d}$. For 
example from the lowest emitter sub-band there cannot be any contribution to the 
tunnelling 
through those dot states with positive angular momenta.

 The overlap integrals~\ref{eq:over} can be calculated analytically and are given 
by~\cite{lan}:
\begin{eqnarray}
\label{eq:overb}
M_{n_{d},m_{d}:n_{e},m_{e}}^{\|}=\\ \nonumber
\delta_{m_{d},m_{e}}\frac{\alpha_{d}\alpha_{e}}{2
}\Gamma(\gamma)
\lambda^{-n_{d}-n_{e}-\gamma}(\lambda-k_{d})^{n_{d}}(\lambda-k_{e})^{n_{e}}  \\	
\nonumber
\times
F(-n_{d},-n_{e},\gamma,\frac{k_{d}k_{e}}{(\lambda-k_{d})(\lambda-k_{e})}) 
\end{eqnarray}
where $\alpha_{d,e}=\frac{1}{a_{d,e}^{\gamma}} \sqrt{\frac{(|m_{d,e}|+n_{d,e})!}
{2^{|m_{d,e}|}n_{d,e}!|m_{d,e}|!^{2}}}$,
$a_{d,e}=$ $\sqrt{\frac{\hbar}{2m^{*}\omega_{d,e}(B)}}$. F is the hypergeometric 
function~\cite{lan},
$\lambda=\frac{1}{4}\frac{a^{2}_{d}+a^{2}_{e}}{a^{2}_{d}\times a^{2}_{e}}$,
$k_{d,e}=\frac{1}{2a^{2}_{d,e}}$, $\gamma=|m_{d,e}|+1$ and 
$\omega_{e}=\omega_{c}/2$.

\begin{figure}[!p]
\centering
\includegraphics[width=4.5in]{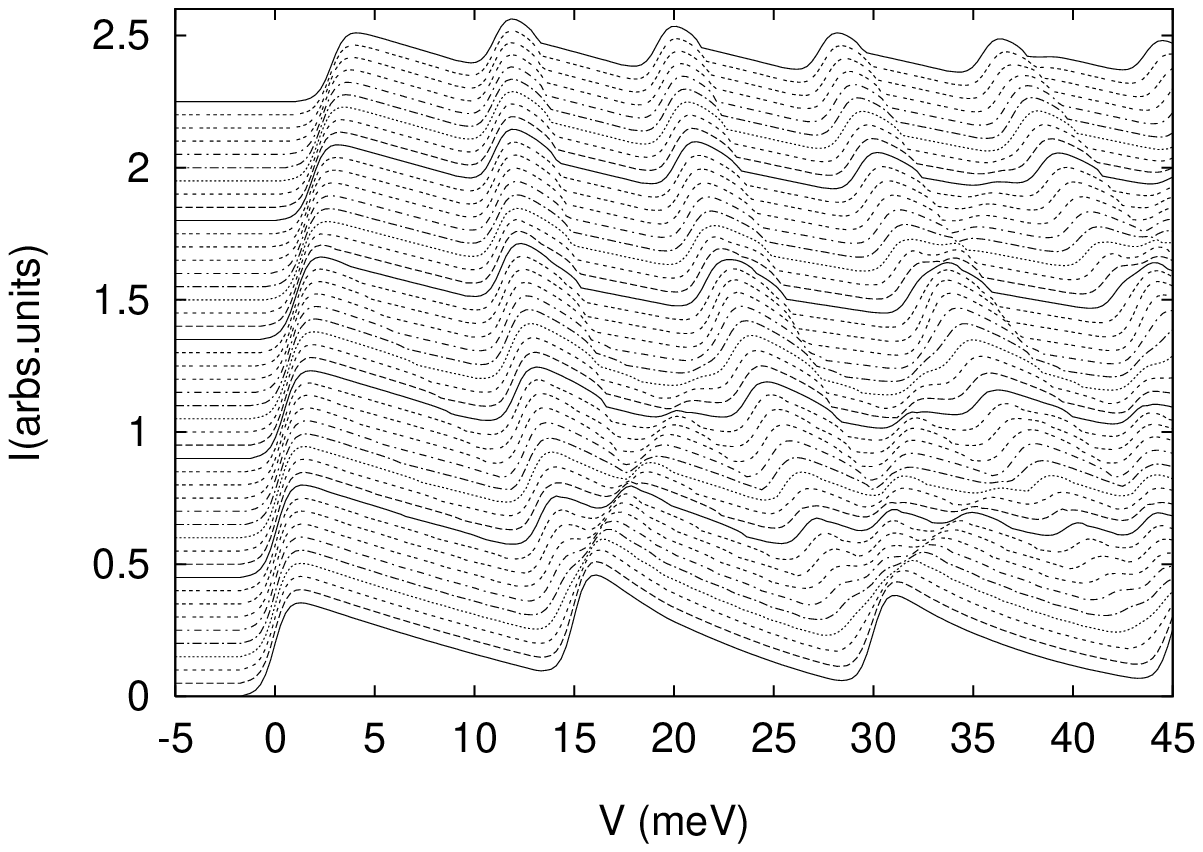}
\caption{I-V for several dot states with a 3D emitter at $T=4K$; V=0 corresponds to the 
threshold voltage. From the bottom to the top curve the magnetic field varies from 
0 to 10.25T with a step of 0.25T. The curves are shifted for clarity.}
\label{fig:iv3D}
\end{figure}

Therefore combining formula~\ref{eq:overb} with the overlap integrals $M^{\bot}$ 
in the direction of the 
current as outlined by Payne~\cite{pay86} and by Liu and Aers~\cite{liu88}, one 
can
calculate the current flowing through the quantum dot in an analytical way.
The above technique has been adopted to calculate the I-V characteristics of 
3D-0D-3D structures in the presence of an external magnetic field applied in the 
vertical direction, i.e. parallel to the current. Fig.~\ref{fig:iv3D} shows the 
calculated I-V characteristics at $T=4K$ for several values of the magnetic field, 
calculated with 
$\hbar\omega_{0}=15\mathrm{meV}$. This value of the confinement strength together 
with our choice of $Ga_{0.33}Al_{0.67}As$ barriers of widths 8.7nm and of a $GaAs$ 
quantum well 5.1nm width, is in accordance with some of our experimental devices 
and measurements~\cite{jou98}. At low temperature as the voltage bias is raised, 
electrons tunnel towards unoccupied states in the collector so that tunnelling is 
dominated by the emitter barrier.

\begin{figure}[!p]
\centering
\includegraphics[width=4.5in]{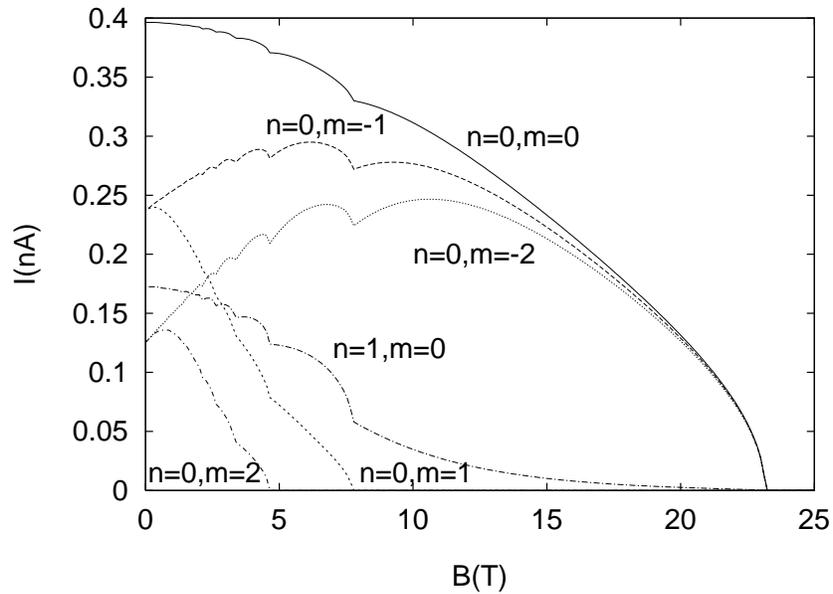}
\caption{Current maximum as a function of B for several dot states in the 3D-0D 
tunnelling case.}
\label{fig:ib3D}
\end{figure}

 The triangular shape of the peaks at B=0T is typical of a 3D-0D tunnelling.
 As expected, the presence of the magnetic field shifts the position of the peaks 
on the voltage scale. This effect is due to the change in the energy of the 
quantum dot states as the magnetic field is increased. More important is the 
current amplitude of the peaks that oscillates as a function of B and have a 
different behaviour from peak to peak. To better illustrate this effect in 
fig.~\ref{fig:ib3D} we show the behaviour of the current amplitude for different 
states. States $n_{d}=0,m_{d}=-2,-1,0,1,2$ and  $n_{d}=1,m_{d}=0$ 
are represented.

The amplitude for all the states shows oscillations as a function of the magnetic 
field. These oscillations have the characteristic $\frac{1}{B}$ dependence and are 
due to the oscillations of the 
density of states at the Fermi energy in the emitter as the magnetic field is 
changed.

Furthermore, as the magnetic field is increased, the peaks show a rather different 
behaviour in 
terms of their amplitude. The peak associated with the ground state shows an 
amplitude that decreases systematically as a function of the magnetic field, while 
the first excited state $n_{d}=0,m_{d}=-1$ initially increases in amplitude, 
attains a maximum, and eventually decreases. The state $n_{d}=0,m_{d}=-2$ has the 
same qualitative features as the $n_{d}=0,m_{d}=-1$ peak, but the maximum is more 
pronounced and shifted at higher magnetic field. These effects have indeed been 
observed recently by B. Jouault et al~\cite{jou98}.

The presence of the magnetic field produces two effects. Firstly it reduces the 
momentum along the direction of the current for the electrons in the contacts by 
moving the sub-bands' minima to higher energies. Secondly it confines the 
electrons 
towards the center of the dot. The first effect tends to diminish the current, 
while the second one increases it by increasing the overlap in the lateral 
direction 
between the quantum dot wavefunctions and those of the contacts. Thus each Landau 
sub-band in the emitter gives a current which attains a maximum at a finite 
magnetic field. The position of this maximum depends on the relative strength of 
the two effects. Moreover increasing the angular momentum $m_{d}$ decreases the 
overlap integral and shifts this maximum to higher fields. The total current is 
obtained by summing the contribution of all the occupied Landau levels in the 
emitter and it follows that the shift of the maximum current peak from the origin 
$B=0$T is a direct measurement of the magnitude of the angular
 momentum $m_{d}$.

Perhaps even more striking is the totally different behaviour of the states 
$n_{d}=0,m_{d}=1$ and $n_{d}=0$,$m_{d}=-1$. At $B=0$T these states are degenerate 
and the radial part of the wave function is identical for both of them for all 
values of the magnetic field. This is confirmed by the fact that at $B=0$ the 
amplitude is the same for both of them. However, the lowest Landau sub-band in the 
contacts does not contain the state $n_{d}=0$,$m_{d}=1$ so that there is no 
contribution to the current flowing through such state from this Landau sub-band. 
Since the radial part of the overlap integrals is at a maximum between states with 
the same principal quantum number, as soon as well defined Landau levels are 
formed in the contacts tunnelling through the state  $n_{d}=0$,$m_{d}=1$ is 
severely 
limited. Indeed whenever the magnetic field is strong enough so that only the 
lowest Landau sub-band is occupied in the contacts, tunnelling through the states 
with angular momentum $m>0$ is altogether forbidden. This is a manifestation of 
the general selection rule previously mentioned, tunnelling through a state with 
positive angular momentum $m$ is forbidden whenever there are only $m$ occupied 
Landau sub-bands in the contacts. The consequence of this is clearly observed in 
fig.~\ref{fig:iv3D}: all the current peaks which are not related to the first 
Landau level in the emitter disappear quickly when the magnetic field increases. 
These peaks related to a positive angular momentum with $n=0$ or to a strictly 
positive value of $n$ (with all values of $m$) shift to higher voltage bias as 
the magnetic field increases. As an example the second current peak due to the 
first excited quantum box state at about 15meV splits into two peaks with $B$ ($n=0$, $m=\pm 1$),
 the third one into three ($n=0$, $m=\pm 2$ and $n=1$, $m=0$)
 and so on. All the current contributions of the peaks 
shifting to higher voltages ($n=0$, $m>0$ or $n>0$) wash out very quickly with the magnetic field.

Finally fig.~\ref{fig:ib3D} shows that the contribution of the first Landau 
level for the state $n_{d}=1,m_{d}=0$ is weak relatively to the other dot states 
$n_{d}=0,m_{d}=0,-1,-2$. This effect reflects the poor overlap of the radial part 
between the states 
$n_{e}=0,m_{e}=0$ and $n_{d}=1,m_{d}=0$. Therefore, the current is sensitive to 
the 
shape of the confining potential, even if the shift of the maximum is a general 
feature for every type of quantum dot.

\section{ Second case: the 2D-0D transport}
The same calculations can be performed for a one-dimensionally confined emitter, 
which corresponds to those experimental structures where an accumulation layer is 
formed in the emitter. In this case a two-dimensional electron gas (2DEG) is 
formed to the left-side of the emitter-dot barrier and the transport properties of 
the 
structures are dominated by the resonant tunnelling between the 2DEG and the 
quantum dot. The form of the confining potential that produces the 2DEG in the 
contacts is roughly of triangular shape and only the lowest level $E_{e}^{0}$ of 
this accumulation layer is usually filled and participate to the tunnelling 
current~\cite{men86}.
 Therefore we are only 
going to consider the lowest emitter state  $E_{e}^{0}$ in the following analysis.
The method adopted in the previous section for the calculation of the I-V 
characteristics can be readily used also in this case. The only difference is that 
the energies and eigenfunctions of the emitter are modified and can be written as:

\begin{equation}
E_{n_{e},m_{e}} =
\hbar\omega_{c}(n_{e}+\frac{m_{e}+|m_{e}|}{2}+\frac{1}{2})+E_{e}^{0}
\end{equation}
the wavefunctions are always given by the equation~\ref{eq:phie} but here 
$\Psi_{e}(z)$ is the same for all the electrons in the accumulation layer. In 
other words the presence of a magnetic field produces a fully-discrete spectrum 
in 
the accumulation layer. Several different forms of the electron wavefunction can 
be taken in 
the triangular groove. However for the scope of our calculations the detail of the 
 wavefunction is not important, the only restriction being that under the barrier 
it attains an exponentially 
decaying form. Furthermore as such wavefunction is the same for all electrons, the 
overlap integral in the growth direction is a constant. Assuming a simple 
Lorentzian spectral function of characteristic width $\Gamma$ for each Landau 
level in the accumulation layer~\cite{nato87}, the current can be calculated as:

\begin{equation}
I=2e|M^{\bot}|^{2}
\sum_{n_{d},m_{d}: n_{e},m_{e}} \! \! \! \!
\frac{
\left|M_{n_{d},m_{d} n_{e},m_{e}}^{\|}\right|^{2}
f(E_{d}-E_{f})
\Gamma/ \pi}{\Delta E_{e,d}^{2}+\Gamma^{2}}
\end{equation}
where $\Delta E_{e,d}$ is the difference in energy between the state in the 
emitter and the state in the dot at a given voltage bias.

\begin{figure}[!p]
\centering
\includegraphics[width=4.5in]{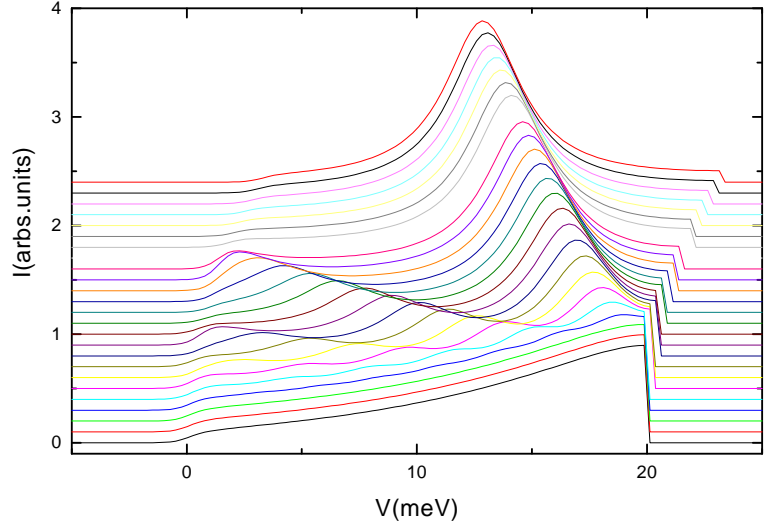}
\caption{I-V curves of the ground state for a 2D emitter at $T=4K$; V=0 corresponds to the 
threshold voltage. From the bottom to the top curve the magnetic field varies from 
0 to 12T with a step of 0.5T.The curves are shifted for clarity. }
\label{fig:iv2D}
\end{figure}

 Figure~\ref{fig:iv2D} shows the contribution to the I-V characteristics for 
the ground state in the dot as the magnetic field increases. At B=0T the ground 
state current increases exponentially with the energy, as expected if there is 
neither magnetic nor geometric lateral confinement in the contacts. Indeed the 
overlap integral is greater with the states of the emitter near the conduction 
band edge because their small lateral momentum is of the same order of magnitude 
than size of the dot state. Thus the sharp closure at V=20meV arises when the 
bottom of the conduction band in the emitter is aligned with the state of the dot. 
The Landau 
levels in the accumulation layer become well-defined as the ratio 
$\Gamma/\omega_{c}<1$, and their formation is observable in the current flowing 
through 
the quantum dot ground state by the presence of several peaks. The Landau levels 
are shifted towards the emitter Fermi level by the magnetic field. This is 
reflected in fig.~\ref{fig:iv2D} by the fact that the peak associated to each 
Landau level shifts towards the threshold voltage as the magnetic field increases.

As the overlap integrals increase with B, the oscillations shown by the current as 
a function of B increase in amplitude with the magnetic field. It is important to 
notice that for a 2D emitter the increase in the amplitude of the current with B 
is much more pronounced than the one seen for 3D contacts. This is due to the fact 
that in 3D the increase in the overlap between the wavefunctions in the lateral 
directions is counteracted by a decrease in the electron momentum along the 
direction of the current, as shown in the previous section. The latter effect does 
not occur in a 2D emitter and the change in the amplitude of the current resonances as 
a function of the field is only 
due to the change in the overlap integral. This is a key difference between 
tunnelling from 2D and 3D contacts. 
The contribution to the current associated with each Landau level is directly 
proportional to the overlap integral between the Landau level and the quantum dot 
wavefunction. As the overlap with the first Landau level is the only one which 
tends to one when B increases, the contribution of the first Landau level is 
dominant.
 
To the best of our knowledge, such an increase has never been observed and only 
3D-0D and 1D-0D
transport has been reported. Nevertheless our theory clearly predict this effect 
in quantitative terms and highlights the profound difference in the behaviour of 
the fine features associated to the quantum dot states as a function of the 
magnetic field depending on whether an accumulation layer is formed or otherwise. 
Furthermore, at least for the quantum dot ground state, our prediction is not 
affected by the presence of the electron-electron interaction in the quantum dot 
as when the ground state becomes resonant there is only one electron in the 
quantum dot.

\begin{figure}[!p]
\centering
\includegraphics[width=4.5in]{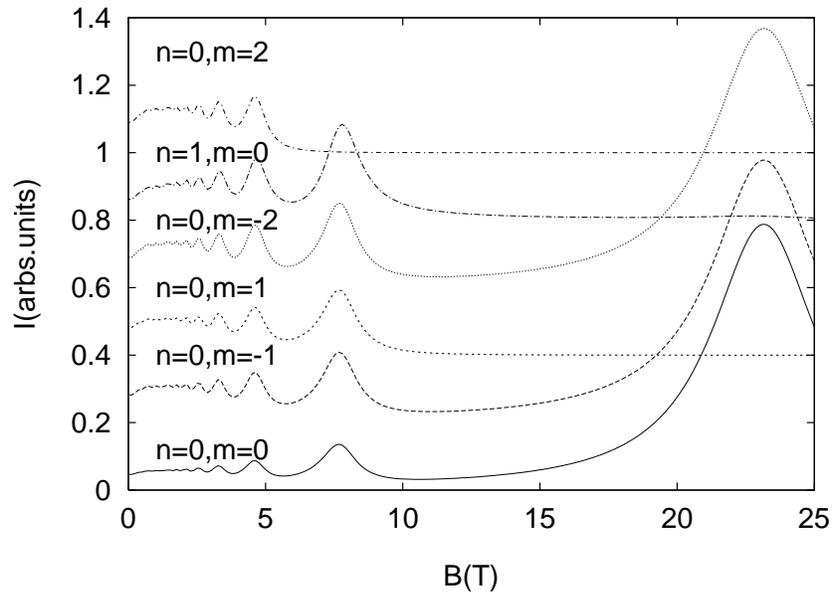}
\caption{Current threshold as a function of B for different dot states in the 
2D-0D tunnelling case.}
\label{fig:ib2D}
\end{figure}

Finally, in fig.~\ref{fig:ib2D} we show the current threshold for different dot 
states as a function of the magnetic field. As in the 3D-0D case (see 
fig.~\ref{fig:ib3D}) we clearly observe here the selection rule over the angular 
momentum which suppresses some peaks for the $n_{d}=0,m_{d} > 0$ states, e.g.: 
peaks at about 23T for the current contribution of the first Landau level through 
the dot states $m_{d}=1$ and $2$ or peak at about 7T for the current contribution 
of the second Landau level through the dot state $m_{d}=2$. Let us point out that 
the current contribution of the first Landau level at 23T through the $n_{d}=1, 
m_{d}=0$ is very weak but not suppressed as expected from the angular momentum 
selection rule. Furthermore the increase in the current peak with magnetic field 
can be observed for all those dot states with non-positive angular momentum and 
$n=0$.

\section{ Conclusion}
We have presented a theoretical model that allows us to calculate 
analytically the current flowing through a quantum dot state as a function of the 
magnetic field and the voltage bias. Calculations have  been performed for 3D-0D, 
2D-0D  tunneling. The results show that the amplitude of the current is 
very sensitive to the angular momentum of the quantum dot state. In particular the 
presence of the magnetic field hinders tunnelling through the states with angular 
momentum co-linear with the magnetic field. Furthermore in the case of tunnelling 
from 2D accumulation layers, the current amplitude associated with certain dot 
states is predicted to increase hugely as an external magnetic field is applied. 
The analysis presented in this work can 
be used to interpret transport experiments in quantum dots and directly 
access details of the wavefunctions of the zero-dimensional states~\cite{jou98}.

\section{ Acknowledgments}
We would like to kindly thank A.~Tagliacozzo, A.~Angelucci, G.~Santoro and V.~Marigliano-Ramaglia
 for fruitful discussions. M.B. acknowledges support from the European Community through 
the ESPRIT project FASEM.


\onecolumn

\end{document}